\documentclass[twocolumn,prb,amsmath,showpacs,amssymb,citeautoscript,floatfix]{revtex4}
\def\etal{{ \it et al. }}
\def\prb{{Phys. Rev. B}}
\def\pl{{Phys. Rev. Lett.}}
\usepackage{epsfig}

\begin{document}
\title{ Exchange interactions and Curie temperatures in Cr-based alloys in Zinc Blende structure: 
volume- and composition-dependence
} 
\author{S. K. Bose}
\affiliation{Physics Department, Brock University, St. Catharines,
Ontario L2S 3A1, CANADA                    } 
\author{J. Kudrnovsk\'{y}}
\affiliation{Institute of Physics, Academy of the Sciences of the Czech Republic, Na Slovance 2,
182 21 Prague 8, Czech Republic }
\begin{abstract}
We present calculations of the exchange interactions and Curie temperatures in Cr-based pnictides and
chalcogenides of the 
form CrX with X=As, Sb, S, Se and Te, and the mixed alloys CrAs$_{50}$X$_{50}$ with X=Sb, S, Se, and Te. 
The calculations are performed for Zinc Blende (ZB) structure for 12 values of the lattice parameter
between 5.44 and 6.62 \AA, appropriate for some typical II-VI and III-V semiconducting substrates.
Electronic structure is calculated via the linear muffin-tin-orbitals (LMTO) method in the atomic 
sphere approximation (ASA), using empty spheres to optimize ASA-related errors. Whenever necessary, 
the results have been verified using the full-potential version of the method, FP-LMTO. The disorder 
effect in the As-sublattice for CrAs$_{50}$X$_{50}$  (X=Sb, S, Se, Te) alloys is taken into account 
via the coherent potential approximation (CPA). Exchange interactions are calculated using the linear 
response method for the ferromagnetic (FM) reference states of the alloys, as well as the disordered local moments (DLM) 
states. These results are then used to estimate the Curie temperature from the low and high temperature 
side of the ferromagnetic/paramagnetic transition. Estimates of the Curie temperature are provided, 
based on the mean field and the more accurate random phase approximations. Dominant antiferromagnetic 
exchange interactions for some low values of the lattice parameter for the FM reference states in CrS, 
CrSe and CrTe prompted us to look for antiferromagnetic (AFM) configurations for these systems with 
energies lower than the corresponding FM and DLM values. Results for a limited number of such AFM 
calculations are discussed, identifying the AFM[111] state as a likely candidate for the ground state 
for these cases.
\end{abstract}
\maketitle 
\section{Introduction} 

Half-metallic ferromagnets with high (room temperature and above) Curie temperatures $T_c$ are  ideal  
for spintronics applications, and as such, much experimental and theoretical\cite{Sato1,Sato4,Sato5}
 effort has been devoted 
in recent years to the designing of and search for such materials. Among these,  Cr-doped 
dilute magnetic semiconductors (DMS)
\cite{Saito,Sato6} or Cr-based alloys and in particular CrAs and CrSb
\cite{Akinaga2000,Li2008,Shirai2003,Akinaga2005,Yamana2004,Kahal2007,Galanakis2003,Pask,Ito,Shi,Zhang} in 
Zinc Blende(ZB) structure have attracted particular attention, not only because of the possibility 
of complete spin polarization of the carriers at the Fermi level, but also for their possible high 
$T_c$. Akinaga \etal \cite{Akinaga2000} were able to grow ZB thin films of CrAs on GaAs (001) 
substrates by molecular beam epitaxy, which showed ferromagnetic behavior at temperatures in excess 
of 400 K and magnetic moments of 3$\mu_B$ per CrAs unit. Theoretical calculations by Akinaga 
\etal \cite{Akinaga2000} and several other theoretical calculations since then 
\cite{Kubler2003,Sanyal2003,Pettifor2003,Shirai2003,Pask,Galanakis2003} have verified the half-metallic 
character of CrAs. The high value of $T_c$ has also been supported by some of these 
studies\cite{Kubler2003,Sanyal2003,Pettifor2003}. Thin films of CrSb grown by  solid-source molecular 
beam epitaxy on GaAs, (Al,Ga)Sb, and GaSb have been found to be of ZB structure and ferromagnetic with $T_c$ 
higher than 400  K\cite{Zhao2001}. 

Galanakis and Mavropoulos\cite{Galanakis2003}, motivated by the successful fabrication of ZB CrAs, 
CrSb and MnAs\cite{Ono2001}, have examined the possibility of half-metallic behavior in ordered ZB 
compounds of transition metals V, Cr and Mn with the '$sp$' elements N, P, As, Sb, S, Se and Te. 
Their theoretical 
study shows that the half-metallic ferromagnetic character of these compounds is preserved 
over a wide range of lattice parameters. They also found that the half-metallic character 
is maintained for the transition element terminated (001) surfaces of these systems. 
Yamana \etal \cite{Yamana2004} have studied the effects of tetragonal distortion on ZB CrAs 
and CrSb and found the half-metallicity to survive large tetragonal distortions. Of course, 
the ground states of many of these compounds in the bulk are known to be different from the ZB 
structure, the most common structure being the hexagonal NiAs-type. Zhao and Zunger\cite{Zunger2005} 
have argued that ZB MnAs, CrAs, CrSb, and CrTe are epitaxially unstable against the NiAs structure, 
and ZB CrSe is epitaxially stable only for lattice constants higher than 6.2 \AA, remaining 
half-metallic at such volumes. They also find that even though the ground state of CrS is ZB, it is
antiferromagnetic at equilibrium lattice parameter, and thus not half-metallic.
These results reveal the challenge experimentalists face in
synthesizing these compounds in ZB structure. However, the possibility remains open that 
such difficulties will be overcome with progress in techniques of film-growth and materials 
preparation in general. Recently, Deng \etal \cite{Deng2006} were successful in increasing the 
thickness of ZB-CrSb films to $\sim$ 3 nm by molecular beam
epitaxy using (In,Ga)As buffer layers, and Li \etal \cite{Li2008} were able to grow 
$\sim$ 4 nm thick ZB-CrSb films on NaCl (100) substrates.

In view of the above situation regarding the state of experimental fabrication of these compounds 
and available theoretical results, it would be appropriate to study the variation of magnetic 
properties, particularly exchange interactions and the Curie temperature, of Cr-based pnictides 
and chalcogenides as a function of the lattice parameter.  Towards this goal, we have carried 
out such calculations for the compounds CrX (X=As, Sb, S, Se and Te) and the mixed alloys 
CrAs$_{50}$X$_{50}$ with X=Sb, S, Se and Te. Essentially we study the effect of  anion doping by 
choosing elements of similar atomic sizes (neighboring elements in the Periodic Table), one of which, namely Sb, 
is isoelectronic to As,  while the others (S,Se,Te) bring one more valence electron to the system. 
The mixed pnictide-chalcogenide systems offer further opportunity to study the effects of anion doping.
The alloying with other 3d transition metals (both magnetic, e.g. Fe or Mn,  or non-magnetic , e.g. V) on cation sublattice  
would also change the carrier concentration and bring about  strong d-disorder which 
can additionally modify the shape of the Fermi surface.  This, however, is not the subject of the present paper.

Almost all theoretical studies on these alloys so far address 
aspects of electronic structure and stability of these alloys only. Although a few theoretical estimates of 
exchange interactions 
and the Curie temperature for CrAs at equilibrium lattice parameter have appeared in the literature,
 a detailed study of the volume dependence of these quantities is missing. For the other alloys, CrSb,
CrS, CrSe and CrTe,  no theoretical results for the exchange interaction, Curie temperature  and their volume 
dependence exist at present. The mixed pnictide-chalcogenide systems offer the possibility of not only creating
these alloys over a larger range of the lattice parameter, but also with a larger variation in the
exchange interactions. This is because at low values of the lattice parameter the dominant Cr-Cr exchange interactions in the
chalcogenides can be antiferromagnetic, while for the pnictides they are ferromagnetic. 
 The pnictide-chalcogenide alloying is  important from the experimental viewpoint 
of stabilizing the ZB structure on a given substrate, via the matching of the lattice parameter of the film 
with that of the latter. Although the present study is confined to the ZB structure only, we hope that it will 
provide some guidance to the experimentalists in their search and growth of materials suitable for spintronic devices. 

\section{Electronic structure}

Electronic and magnetic properties of CrX (X=As, Sb, S, Se and Te) and CrAs$_{50}$X$_{50}$ 
(X= Sb, S, Se and Te) were calculated for lattice parameters varying between 5.45 and 6.6 \AA, 
appropriate for some typical II-VI and III-V semiconducting substrates. Calculations were performed 
using the TB-LMTO-CPA method\cite{Kudrnovsky1990,Turek97} and the exchange-correlation potential 
given by Vosko, Wilk and Nusair\cite{VWN}. In our LMTO calculation we optimize the ASA (atomic sphere 
approximation) errors by including empty spheres in the unit cell. We use the fcc unit cell, with 
Cr and X (As, Sb, S, Se and Te) atoms located at (0,0,0) and (0.25,0.25,0.25), respectively, and empty 
spheres at locations (0.5,0.5,0.5) and (-0.25,-0.25,-0.25). For several cases, we have checked the accuracy 
of the LMTO-ASA electronic structures against the full-potential LMTO results\cite{savrasov} and 
found them to be satisfactory. For the mixed alloys CrAs$_{50}$X$_{50}$ (X= Sb, S, Se and Te), the As-sublattice 
of the ZB CrAs structure is assumed to be randomly occupied by equal concentration of As and X atoms.
The disorder in this sublattice is treated under the coherent potential approximation (CPA)\cite{Kudrnovsky1990,Turek97}.

 Our spin-polarized calculations assume a collinear magnetic model. In the following we will present 
 results referred to as FM and DLM.  The FM results follow from the usual spin-polarized 
calculations, where self-consistency of charge- and spin-density yields a nonzero magnetization per 
unit cell. Although we call this the FM result, our procedure does not guarantee that the true ground 
state of the system is ferromagnetic, with the magnetic moments of all the unit cells perfectly 
aligned. This is because we have not explored non-collinear magnetic states, nor all antiferromagnetic (AFM)
states attainable within the collinear model. Indeed, our results for the exchange interactions in 
some cases do suggest the ground states being of AFM or complex magnetic nature. 
For lack of a suitable label, we refer to all spin-polarized calculations giving a nonzero local 
moment as FM state calculations. Within the Stoner model, a nonmagnetic state above the Curie 
temperature $T_c$ would be characterized by the vanishing of the local moments in magnitude. It is 
well-known and universally accepted that the neglect of the transversal spin fluctuations in the 
Stoner model leads to an unphysical picture of the nonmagnetic state and a gross overestimate of $T_c$. 
An alternate description of the nonmagnetic state is provided by the disordered local moment (DLM) model, 
where the local moments remain nonzero in magnitude above $T_c$, but disorder in magnitude as well as 
their direction above $T_c$ causes the global magnetic moment to vanish. Combining aspects of the Stoner 
model and an itinerant Heisenberg-like model, Heine and co-workers\cite{Heine1,Heine2} have developed a 
suitable criterion for a DLM state to be a more appropriate description of the nonmagnetic state 
than what is given by the Stoner model. Within the collinear magnetic model, where all local axes of 
spin-quantization point in the same direction, DLM can be treated as a binary alloy problem and thus 
described using the coherent potential approximation (CPA)\cite{Hasegawa,Pettifor,Staunton,Pindor}. 
We have carried out such DLM calculations, assuming the Cr-sublattice to be randomly occupied by an 
equal number of Cr atoms with oppositely directed magnetic moments. The object for performing the 
DLM calculations is two-fold. If the total energy in a DLM calculation is lower than the corresponding 
FM calculation, we can safely assume that the ground state (for the given lattice parameter and structure) 
is not FM, albeit of unknown magnetic structure. The exchange interactions in the DLM state can also be 
used to compute estimates of $T_c$, and such estimates of $T_c$ may  be considered as estimates from 
above the magnetic-nonmagnetic transition. $T_c$ computed from exchange interactions in the FM state are 
estimates from below the transition. Of course, if the ground state is known to be ferromagnetic, then 
estimates of $T_c$ based on exchange interactions in the FM reference state are the appropriate ones to consider.

In some cases where the FM results point to the possibility of the ground state magnetic structure being AFM or 
of complex nature, we have carried out a limited number of AFM calculations to provide some insight into this 
problem (see section \ref{subsec:stabilityJq}).

\begin{figure}
\includegraphics[angle=270,width=3.75in]{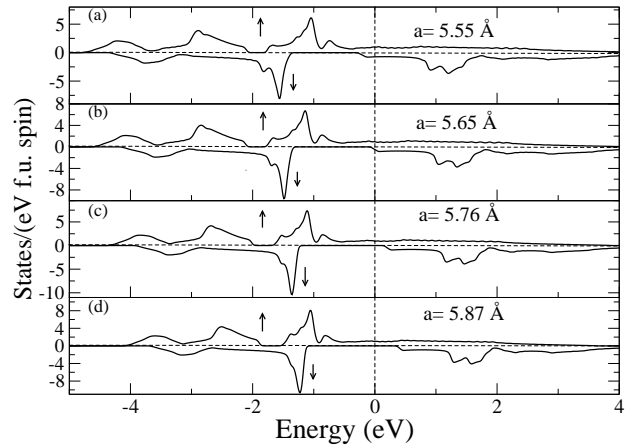}
\caption[]{Spin-resolved densities of states in ZB CrAs$_{50}$Sb$_{50}$ for lattice parameters  (a) 5.55 \AA, (b) 5.65 \AA,
(c) 5.76 \AA$\;$ and 
(d) 5.87 \AA, respectively. }
\label{fig1} 
\end{figure}

\begin{figure}
\includegraphics[angle=270,width=3.75in]{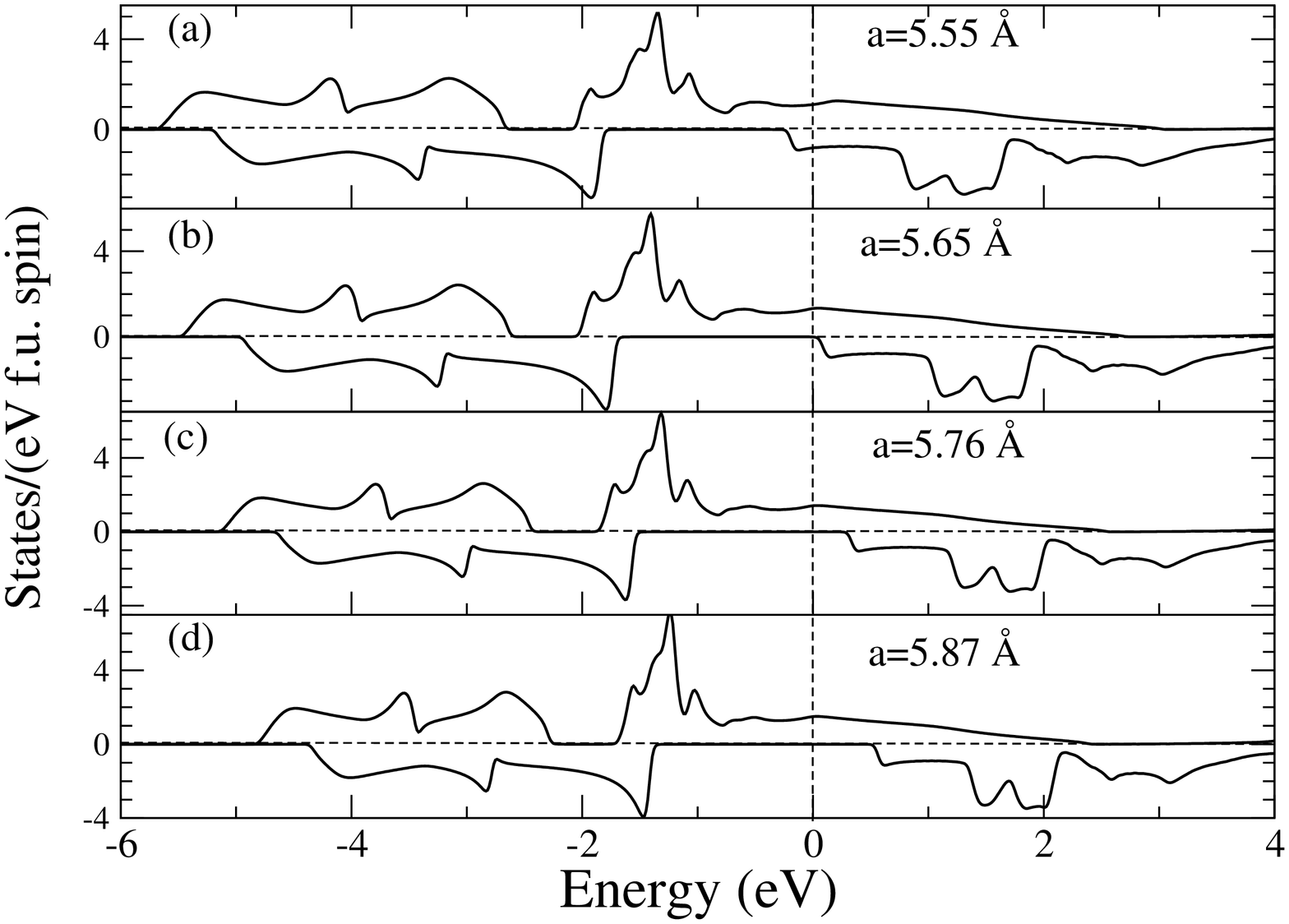}
\caption[]{Spin-resolved densities of states in ZB CrAs$_{50}$Se$_{50}$ for lattice parameters  (a) 5.55 \AA, (b) 5.65 \AA,
(c) 5.76 \AA$\;$ and 
(d) 5.87 \AA, respectively. }
\label{fig2} 
\end{figure}

We have computed the spin-resolved densities of states (DOS) for all the alloys for varying lattice 
parameters, and for both the FM and DLM configurations. The FM calculations show half-metallic character, due to the
formation of bonding and antibonding states involving the $t_{2g}$ orbitals of the Cr-atoms and the  $sp$ orbitals
of the neighboring pnictogen (As, Sb) or chalcogen (S, Se, Te). The hybridization gap is different and takes place in
different energy regions in the two spin channels. The critical values of the lattice parameters above 
which the FM calculations show half-metallic character agree well with those reported by 
Galanakis and Mavropoulos\cite{Galanakis2003}. The DOS for the alloys of the type CrX (X=Sb, S, Se, Te) 
have been presented by several other authors\cite{Galanakis2003,Shi,Zhang} and thus will not be shown 
here. In Figs.\ref{fig1} and \ref{fig2} we 
show the DOS for  the mixed alloys CrAs$_{50}$Sb$_{50}$ and CrAs$_{50}$Se$_{50}$, for lattice parameters above 
and below the critical values for the half-metallic character.
According to Galanakis and Mavropoulos\cite{Galanakis2003}, half-metallicity in ZB CrAs appears between the
lattice parameters of 5.45 and 5.65 \AA.  The latter corresponds to the lattice parameter of the GaAs substrate.
  For CrSb half-metallicity
appears at a lattice parameter between 5.65 and 5.87 \AA. The mixed alloy CrAs$_{50}$Sb$_{50}$,
as shown in Fig.\ref{fig1}, is not quite half-metallic at the lattice parameter of 5.65 \AA, and fully half-metallic
at the lattice parameter of 5.76 \AA. Replacing Sb with Se in the above alloy, i.e. for CrAs$_{50}$Se$_{50}$,
brings the critical lattice parameter down slightly. As shown in Fig.\ref{fig2}, at a lattice parameter of 5.65 \AA,
CrAs$_{50}$Se$_{50}$ is half-metallic, although barely so. In our calculation CrS and CrSe are  half-metallic at a
lattice parameter of 5.65 \AA, and not so at a lattice parameter of 5.55 \AA. CrTe is not half-metallic at a lattice parameter 
of 5.76 \AA, but at a lattice parameter of 5.87 \AA. For both CrS and CrSe the critical value should be close to 5.65 \AA, 
and for CrTe it should be close to 5.87 \AA.

Note that in general the half-metallic gap is larger in the chalcogenides than in the pnictides. This is due to larger Cr-moment
(see section \ref{sec:mag.mom.}) for the chalcogenides, which results in larger exchange splitting. This explains the difference
in the half-metallic gaps in Figs. \ref{fig1} and \ref{fig2} for similar lattice parameters.

\begin{figure}
\includegraphics[angle=270,width=3.75in]{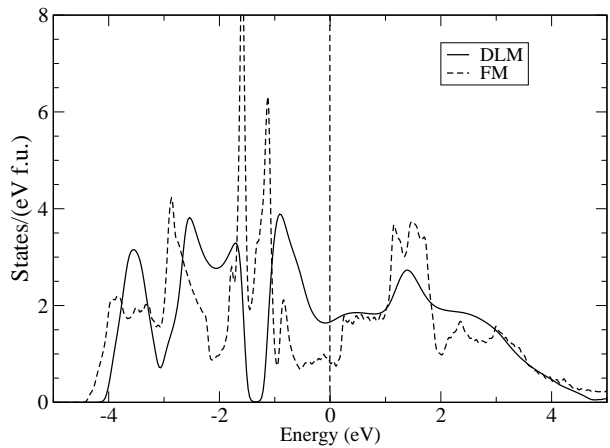}
\caption[]{A comparison of the total densities of states in ZB CrAs with lattice parameter 5.65 \AA$\;$ for 
the DLM and  FM states.}
\label{fig3} 
\end{figure}

Fig. \ref{fig3} compares the total DOS of CrAs for the FM and DLM calculations for the equilibrium 
lattice parameter  5.65 \AA. Higher DOS at the Fermi level for the DLM calculation, compared with 
the FM calculation, is an indication that the band energy is lower in the FM state. Indeed, as 
indicated in Table \ref{table1}, compared with the DLM state the total energy for ZB CrAs is lower 
in the FM state for the lattice parameters from 5.44 to 5.98 \AA. In fact, this holds for lattice 
parameters up to 6.62 \AA, showing the robustness of ferromagnetism in CrAs over  a wide range 
of the lattice parameter. This is also true for CrSb.

\begin{figure}
\includegraphics[angle=270,width=3.75in]{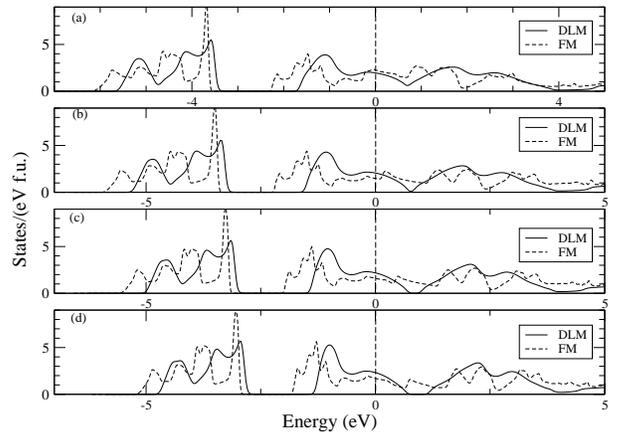}
\caption[]{Total densities of states in CrS in the DLM and FM states for lattice parameters 
(a) 5.44 \AA, (b) 5.55 \AA, (c) 5.65 \AA, and (d) 5.76 \AA, respectively.}
\label{fig4} 
\end{figure}

In Table \ref{table1} we show the variation of total energies per atom in Ry  with the lattice parameter for 
CrX (X=As, Sb, S, Se and Te) in the DLM and FM states. The purpose of tabulating these energies is not 
to determine the bulk equilibrium lattice parameters in the ZB structure, as this has already been done 
by several authors\cite{Galanakis2003,Shi,Zhang}. Our results for equilibrium ZB phase lattice parameters 
agree with those found by Galanakis and Mavropoulos\cite{Galanakis2003}. The important point is that for 
CrS and CrSe at low values of lattice parameters the DLM energies are lower than the FM energies, showing 
clearly that the FM configuration is unstable. The result for CrS is in line with the observation by 
Zhao and Zunger\cite{Zunger2005}, who find ZB CrS to be antiferromagnetic with an equilibrium lattice 
parameter of 5.37 \AA. As shown later (section \ref{sec:exchange}), the exchange 
coupling constants for the Cr atoms in the FM calculations are negative, indicating the 
instability of the ferromagnetic spin 
alignment. The tendency to antiferromagnetism in CrSe at compressed lattice 
parameters is also revealed  in a study by Sasaio$\tilde{g}$lu \etal\cite{Sasaiolglu} For CrTe at lower 
lattice parameters the FM energy is lower than the DLM energy, but the exchange constants for the Cr-atoms in the
FM calculations are still 
negative (see discussion in section \ref{sec:exchange}), signaling the possibility of the ground states in CrTe
 at low values of the lattice parameter being neither DLM nor FM. Note that in our discussion  
ground state implies the lowest energy state in ZB structure.  For CrS, CrSe and CrTe the 
ground states at low lattice parameters can be of an antiferromagnetic (AFM) or  complex magnetic structure. 
A lower total energy may also mean a lower band energy, and in some cases, the latter may be reflected 
in a lower density of states at the Fermi level. This is shown in Fig.\ref{fig4}, where for CrS at 
the lowest lattice parameter of 5.44 \AA$\;$ the DOS at the Fermi level is lower in the DLM state than 
in the FM state. The deviation from ferromagnetism at low values of the lattice parameter for CrS, CrSe and CrTe is 
also revealed by our study of the lattice Fourier transform of the exchange interaction between the 
Cr atoms in the FM state (section \ref{sec:exchange}).  The 
search for an antiferromagnetic state with lower energy is possible within our collinear magnetic 
model by enlarging the unit cell in various ways. We have pursued this issue to a limited extent, 
by considering 001, 111 AFM configurations for CrS, CrSe and CrTe at low values of the lattice parameter
(see discussion in section\ref{sec:exchange}). A satisfactory 
resolution of such issues is possible only by going beyond the collinear model.
\begin{table*}
\caption{Comparison of total energies in the FM and DLM states as a function of the lattice parameter 
for CrAs, CrSb, CrS, CrSe, and CrTe. Results for six lattice parameter values are shown. Calculations include six additional lattice 
parameters beyond 5.98\AA, reaching a maximum of 6.62 \AA. For all these additional lattice parameters FM energy is always lower
than the corresponding DLM energy, indicating that ferromagnetism is favored at higher lattice parameters.}
\label{table1}
\begin{ruledtabular}
\begin{tabular}{lcccccc}
Lattice parameter (\AA) & 5.44 & 5.55 & 5.65 & 5.76 & 5.87& 5.98  \\
{\bf CrAs}\\ 
DLM energy & -1653.4039 & -1653.4021& -1653.3995 &  -1653.3960 & -1653.3920 & -1653.3876 \\
FM energy & -1653.4068 & -1653.4055 & -1653.4034 & -1653.40026 & -1653.3966 & -1653.3922 \\
{\bf CrSb}\\ 
DLM energy & -3762.4738 &  -3762.4781 & -3762.4807 & -3762.4821 & -3762.4822 & -3762.4814\\
FM energy & -3762.4770 & -3762.4813 & -3762.4841 & -3762.4857 & -3762.4862 &-3762.4856\\ 
\\
\hline\\
{\bf CrS}\\
DLM energy & -723.4504 & -723.4468 & -723.4426 & -723.4381 & -723.4332 &-723.4280\\
FM energy & -723.4491 &  -723.4464 & -723.4434 & -723.4395 & -723.4351 & -723.4302\\
{\bf CrSe}\\ 
DLM energy &  -1737.9146 & -1737.9139 & -1737.9123 &-1737.9100 &  -1737.9071 & -1737.9036\\
FM energy & -1737.9139 & -1737.9132 & -1737.9124 & -1737.9110 & -1737.9087 & -1737.9057 \\
{\bf CrTe}\\ 
DLM energy & -3918.9074 & -3918.9124 & -3918.9158 & -3918.9179 & -3918.9189 & -3918.9188\\
FM energy & -3918.9078 & -3918.9128 & -3918.9161 & -3918.9181 & -3918.9194 & -3918.9201\\
\end{tabular}
\end{ruledtabular}
\end{table*}
\section{Magnetic moments}
\label{sec:mag.mom.}
\begin{figure}
\includegraphics[angle=270,width=3.75in]{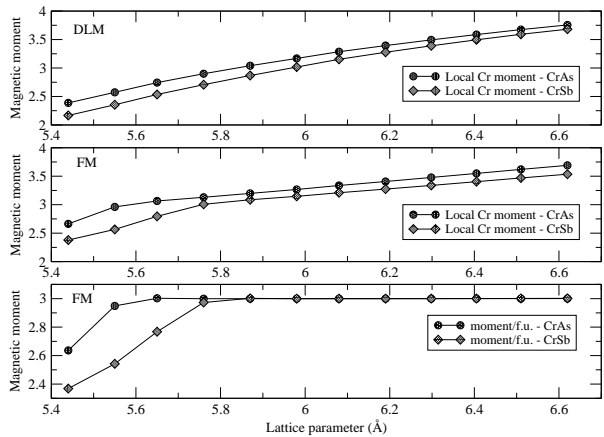}
\caption[]{Magnetic moment in ZB CrAS and CrSb as a function of lattice parameter in the 
FM and DLM  states. In all cases studied (Figs. \ref{fig5}-\ref{fig8}), FM calculations produce 'induced
moments' on non-Cr spheres representing the X-atoms (X=As, Sb, S, Se, Te), and one set of empty spheres. The DLM 
calculations produce no such 'induced moments', i.e., the moments reside on the Cr-atoms only. See text for discussion.}
\label{fig5}
\end{figure}

\begin{figure}
\includegraphics[angle=270,width=3.75in]{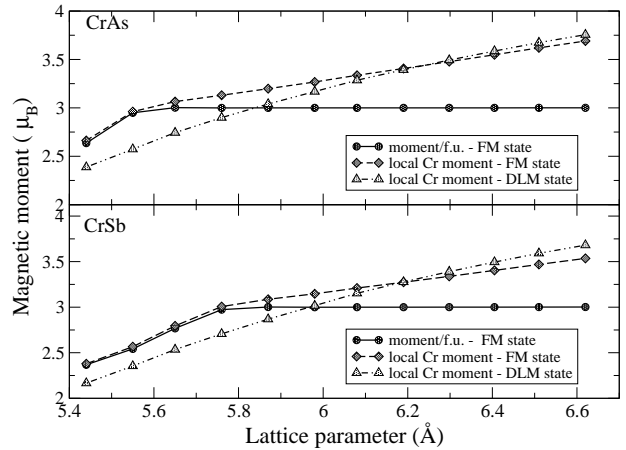}
\caption[]{A comparison of the moment/f.u. with the local Cr moment in the FM state as well 
as the DLM state for ZB CrAs (upper panel) and CrSb (lower panel).  }
\label{fig6}
\end{figure}

Our spin-polarized calculations for the FM reference states lead to local moments not only on the Cr 
atoms, but also on  the other atoms (As, Sb, S, Se, and Te) as well as the empty spheres. Sandratskii 
\etal\cite{Sandratskii} have discussed the problem associated with such 'induced moments' 
in case of the Heusler alloy NiMnSb and the hexagonal phase of MnAs. Usually such systems can 
be divided into sublattices with robust magnetic moments and sublattices 
where moment is induced under the influence of the former. These authors argue that 
the treatment of the induced moments as independent variables in  a Heisenberg Hamiltonian 
may lead to artificial features in the spin-wave spectra, but these artificial features do 
not drastically affect the calculated Curie temperatures of the two alloys, NiMnSb and
hexagonal MnAs. Clearly, in our case the sublattice with the robust magnetic moment is the
Cr-sublattice. Among the three other sublattices, the magnitudes of the induced moments
decrease in the following order for the two robust ferromagnets CrAs and CrSb: X-sublattice (X=As, Sb), 
sublattice ES-1 (the sublattice of empty spheres that is at the same distance with respect to the Cr-sublattice
 as the X-sublattice ),
sublattice ES-2 (sublattice of empty spheres further away from the Cr-sublattice). This trend
is particularly valid for low values of the lattice parameter. 
The induced moments originate from the tails of the orbitals (primarily $d$) on the nearby
Cr-atoms. This is particularly true for the moments induced on the empty spheres. 
The magnitudes of the induced 
moments on the two empty sphere sublattices decrease as the lattice parameter increases, and so
do the differences in their magnitudes. The signs of the moments on ES-1 and ES-2 may be the
same for small lattice parameters, but are opposite for large lattice parameters. The sign
of the moment on the X-sublattice is opposite to that on the Cr-sublattice and the magnitudes
of the moments on the two sublattices increase with increasing lattice parameters, due to
decreased hybridization between Cr-$d$ and X-$sp$ orbitals. Above a critical value of the
lattice parameter, the moment per formula unit (f.u.) saturates at a value of 3.0 $\mu_B$,
as the half-metallic state is achieved,
while the local moments on the Cr- and X-sublattices increase in magnitude, remaining opposite in sign.
The maximum ratio between the induced moment on X (X=As,Sb) and the moment on Cr is 0.18 for CrAs
and 0.15 for CrSb, occurring at the highest lattice parameter of 6.62 \AA$\;$ studied. The maximum
ratio between the induced moment on ES-1 and that on Cr is 0.06, occurring at the lowest lattice parameter
of 5.44 \AA$\;$ studied.
Magnetic moments of CrAs and CrSb per formula unit (f.u.) as well as the local moment at the 
Cr site are shown in Fig. \ref{fig5},  where we compare the two compounds with each other 
for their magnetic moments in the FM and DLM states. The same results are presented in Fig. 
\ref{fig6}, comparing the moment per f.u.in the FM state with the Cr local moment in the FM 
and DLM states separately for each compound. It is to be noted that there are no induced moments
for the DLM reference states, i.e. the moments on the non-CR sublattices are several orders of magnitude 
smaller than the robust moment
on the Cr atoms. The total moment per formula unit in the DLM 
state is zero by construction. The local moment on the Cr atom for the DLM reference state is usually less than the 
corresponding FM value for
smaller lattice parameters, and larger for larger lattice parameters (Fig.\ref{fig6}).

\begin{figure}
\includegraphics[angle=270,width=3.75in]{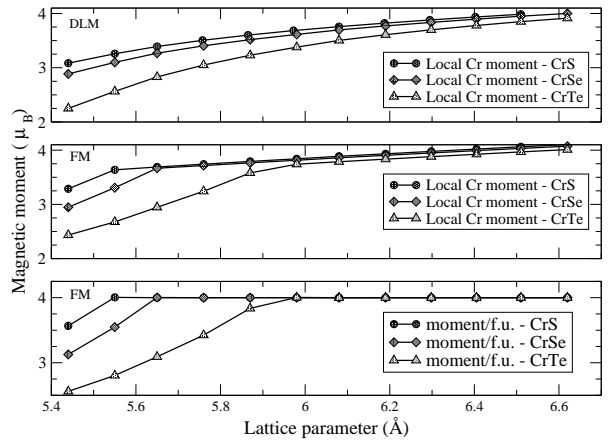}
\caption[]{Magnetic moments in ZB CrS, CrSe and CrTe as a function of lattice parameter 
in the FM and DLM ground states.   }
\label{fig7}
\end{figure}

\begin{figure}
\includegraphics[angle=270,width=3.75in]{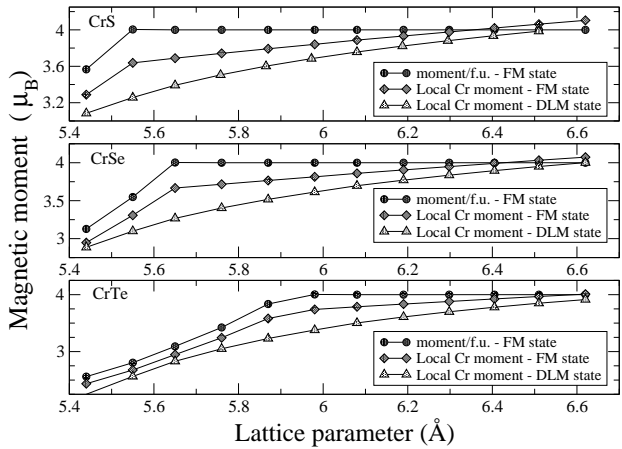}
\caption[]{A comparison of the moment/f.u. with the local Cr moment in the FM state as 
well as the DLM state for ZB CrS(top panel), CrSe (middle panel) and CrTe (bottom  panel).     }
\label{fig8}
\end{figure}

Similar trends in the variation of the local moment on Cr and the induced moments on the other sublattices for the
FM reference states as a function of lattice parameter are revealed for CrX (X=S, Se, Te), except that the
moments on ES-1 are always an order of magnitude larger than those on ES-2. I addition, the induced moments on
ES-2 are $\sim 6-10$ times larger than those on X sublattice for smaller values of the lattice parameter, with the two
becoming comparable in magnitude for larger lattice parameters. The induced moments on ES-1 and X-sublattices are never
larger than $\sim$ 5\% of the moment on the Cr atoms. The induced moments for the DLM reference states are
several orders of magnitude smaller than the Cr-moments, and can be safely assumed to be zero. Results for ZB CrS, CrSe and CrTe are 
presented in Figs.\ref{fig7} and \ref{fig8}.   The 
moment per f.u. reaches the saturation values of  4$\mu_B$ for 
CrS, CrSe, and  CrTe in the half-metallic state, as discussed in detail by Galanakis and 
Mavropoulos\cite{Galanakis2003}. The saturation values of the moments for all these alloys (CrX, X=As, Sb, S, Se, and Te)
 satisfy the so-called 
``rule of 8'': $M=\left(Z_{tot}-8\right)\mu_{B}$, where $Z_{tot}$ is the total number of valence 
electrons in the unit cell. The number 8 accounts for the fact that in the half-metallic state 
the bonding $p$ bands are full, accommodating 6 electrons and so is the low-lying band formed 
of the $s$ electrons from the $sp$ atom, accommodating 2 electrons.  The magnetic moment then 
comes from the remaining electrons filling the $d$ states, first the $e_g$ states and then the 
$t_{2g}$. The saturation value of 3$\mu_B$/f.u., or the half-metallic state, appears for a 
larger critical lattice constant in CrSb than in CrAs. Similarly, the critical lattice constants 
for the saturation magnetic moment of 4$\mu_B$/f.u. are in increasing order for CrS, CrSe and 
CrTe. The local moment on the Cr atom can be less/more than the saturation value,
depending on the moment induced on the non-Cr atoms and empty spheres. 

\begin{figure*}
\includegraphics[angle=270,width=6in]{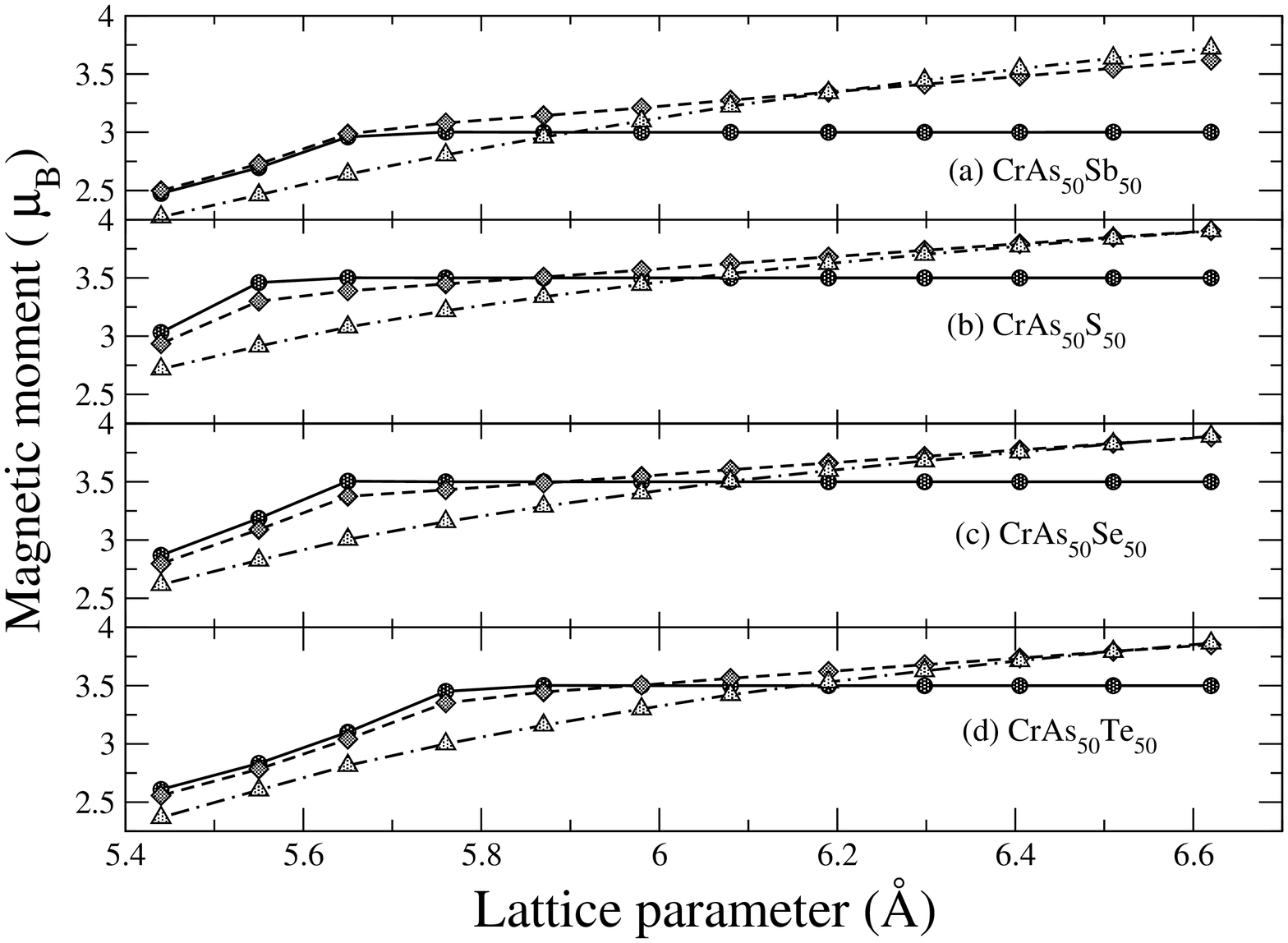}
\caption[]{A comparison of the moment/f.u. with the local Cr moment in the FM state as well as the DLM state for ZB 
(a) CrAs$_{50}$Sb$_{50}$, (b) CrAs$_{50}$S$_{50}$, (c) CrAs$_{50}$Se$_{50}$, and (d)CrAs$_{50}$Te$_{50}$. Circle: 
moment/f.u.- FM state, diamond: local Cr-moment- FM state, triangle: local Cr-moment- DLM state.      }
\label{fig9}
\end{figure*}

Fig.\ref{fig9} shows the variation of the magnetic moment with the lattice parameter for the random alloys CrAs$_{50}$X$_{50}$
(X=Sb, S, Se, Te), where 50\% of the As-sublattice is randomly occupied by X-atoms. The saturation moment per f.u. for
CrAs$_{50}$Sb$_{50}$ in the half-metallic state is 3$\mu_B$, with the results falling between those for CrAs and CrSb 
shown in Fig.\ref{fig6}.  For CrAs$_{50}$X$_{50}$ (X=S, Se, Te), the saturation moment per f.u. is 3.5$\mu_B$. The 
local Cr-moment deviates from the saturation value in the half-metallic state, being higher than the saturation 
value for all lattice parameters above 6.1\AA.

From Figs.\ref{fig5}-\ref{fig9} it is clear that the magnetic moment per formula unit is closer to the magnetic moment of the Cr 
atoms in the FM calculations than in the DLM calculations. Local Cr-moments in the DLM calculations are 
suppressed w.r.t. the FM results for low lattice parameters and enhanced for larger lattice parameters.  As shown in TABLE \ref{table1} the
total energy of the FM state is lower than that of the corresponding DLM state in almost all cases, except
for some compressed lattice parameters for CrS and CrSe.  However, the consideration of the DLM state does 
provide an advantage in that there are no associated induced moments, i.e., the DLM calculations produce moments 
that reside on the robust 
magnetic sublattice only. Mapping of the total energy on to a Heisenberg Hamiltonian, therefore, does not result in 
exchange interactions involving atoms/spheres with 'induced moments' and all associated artificial/non-physical features 
referred to by Sandratskii \etal\cite{Sandratskii} 

\section{Exchange interaction and Curie temperature}
\label{sec:exchange}
\subsection{Mapping onto a Heisenberg Hamiltonian and related issues}
Currently, most \textit {first-principles} studies of the thermodynamic properties of itinerant magnetic systems 
 proceed via mapping \cite{Sandratskii,Pajda2001} the system energy onto a classical Heisenberg model:
\begin{equation}\label{e1}
H_\mathrm{eff} = - \sum_{i,j} \, J_{ij} \,
{\bf e}_{i} \cdot {\bf e}_{j} \ ,
\end{equation}
where $i,j$ are site indices, ${\bf e}_{i}$ is the unit vector
pointing along the direction of the local magnetic moment at
site $i$, and $J_{ij}$ is the exchange interaction between the moments at
sites $i$ and $j$. The validity of this procedure is justified on the basis of  the adiabatic hypothesis- the assumption that 
the magnetic moment directions  are 'slow variables' on all the characteristic electronic time scales relevant to the problem, 
and thus can be treated as classical parameters. The energy of the system for a given set of magnetic moment directions is 
usually calculated via methods based on
density functional theory (DFT). 

One of the most widely used mapping procedures is due to Liechtenstein 
\etal \cite{Liechtenstein84I,Liechtenstein87,Liechtenstein84II,Liechtenstein88,Gubanov92} It 
involves writing the change in the energy due to the deviation of a single spin from a reference state 
in an  analytic form using the multiple scattering formalism and by 
appealing to the magnetic variant of the Andersen force 
theorem\cite{Andersen,Heine}.  The force theorem, derived originally for the change of total energy due to
a deformation in a solid, dictates that the differences in the energies of various magnetic configurations can be 
approximated by the differences in the band energies alone\cite{Liechtenstein84I,Liechtenstein87,Oswald}. The energy of 
a magnetic excitation related to the
rotation of a local spin-quantization direction can be calculated from the spinor rotation of the
ground state potential. No self-consistent calculation for the excited state is necessary. A second approach is
based on the total energy calculations for a set of collinear magnetic structures, and extracting the exchange parameters 
by mapping the total energies to those coming from the Heisenberg model given by Eq.(\ref{e1}). Such calculations can be done 
using any of the standard DFT methods. However, unlike the magnetic force theorem method, where the exchange interactions can be 
calculated directly for a given structure and between any two sites, several hypothetical magnetic configurations and sometimes 
large supercells need to be considered to obtain the values of a modest number of exchange interactions. In addition, 
some aspects of environment-dependence of exchange interactions are often simply ignored.
The  difference between these two approaches is, in essence, the same as that between the generalized 
perturbation method (GPM)\cite{Ducastelle,Sluiter} and the 
Connolly-Williams method\cite{C-W,Lu} in determining the effective pair interactions in ordered and disordered alloys. 
A third approach is a variant of the second approach, where the energies of the system in various magnetic configurations 
corresponding to spin-waves of different wave-vectors are calculated by employing the generalized Bloch theorem for spin-spirals\cite{Sandratskii91}. The inter-atomic exchange interactions can be calculated by equating these energies 
to the Fourier transforms of the classical Heisenberg-model energies.   This approach, known as the 'frozen magnon approach', 
is similar to the 'frozen phonon approach' for the study of lattice vibrations in solids.

In this work, we have used the method of Liechtenstein \etal,
which was later implemented for random  magnetic systems by Turek \etal, using CPA and the TB-LMTO method\cite{Turek2006}. 
The exchange integral in Eq.(\ref{e1}) is given by
\begin{equation}\label{eq-Jij}
 J_{ij} = \frac{1}{4\pi} \lim_{\epsilon\rightarrow 0^+} Im \int tr_L \left[\Delta_i(z)g_{ij}^{\uparrow}(z)
\Delta_j(z)g_{ji}^{\downarrow}
\right] dz \; ,
\end{equation}
where $z=E+i\epsilon$ represents the complex energy variable, $L=(l,m)$, and $\Delta_i(z)= P_i^{\uparrow}
(z)-P_i^{\downarrow}(z)$, representing the difference in the potential
functions for the up and down spin electrons at site '$i$'. In the present work 
$g_{ij}^{\sigma}(z) (\sigma=\uparrow,\downarrow)$ 
represents the matrix elements of the Green's function of the medium for the up and down spin electrons. 
For sublattices with disorder, this is a configurationally averaged Green's function, obtained via using the prescription of CPA.  
The integral in this work is performed in the complex energy plane, 
where the contour includes the Fermi energy $E_F$. The quantity $J_{ij}$ given by Eq. (\ref{eq-Jij}) 
includes direct-, indirect-, double-exchange and superexchange interactions, which are often treated 
separately in model calculations. The negative sign in Eq.(\ref{e1}) implies that positive and negative values of $J_{ij}$  
are to be interpreted as representing ferromagnetic and antiferromagnetic interactions, respectively.

A problem with the mapping of the total energy to a classical Heisenberg Hamiltonian following the approach
of Liechtenstein \etal \cite{Liechtenstein84I,Liechtenstein87,Liechtenstein84II,Liechtenstein88,Gubanov92} is that it generates
exchange interactions between sites, where one or both may carry induced moment(s). Of course this
problematic scenario appears only for the FM reference states, as the DLM reference states do not generate
induced moments. In the present work the Liechtenstein mapping procedure, applied to FM reference sates,
generates exchange interactions  between the Cr atoms, between Cr and other atoms X (X=As, Sb, S, Se, Te),
and also between Cr atoms and the nearest empty spheres ES-1. Depending on the lattice parameter, this latter 
interaction is either stronger than or at least comparable to that for the Cr-X pairs. The exchange interactions between 
Cr atoms and the furthest empty spheres
ES-2 are always about one or two orders of magnitude smaller than the Cr-ES1 interactions and can be neglected.
In CrAs, the ratio of the nearest neighbor Cr-ES1 to Cr-Cr interaction varies from 0.2-0.25 at low lattice parameters to
0.06-0.07 at high values of the lattice parameter. In CrSb, these ratios are smaller, varying between 0.14 and 0.05.
The Cr-ES1 exchange interactions are also relatively strong in magnitude  in CrS, CrSe and CrTe. One important point is
that while these interactions are positive for nearest neighbors for all lattice parameters, Cr-Cr nearest neighbor interaction
is negative for low values of lattice parameters in CrS and CrSe. In CrTe, this interaction changes sign from positive to negative and then
back, as the lattice parameter is varied in the range 5.44-6.62 \AA. As mentioned earlier, the calculation for the DLM reference
states do not produce induced moments, and thus no exchange interactions other than those between the Cr atoms.

Sandratskii\cite{Sandratskii} \etal have discussed the case when, in addition to
the interaction between the strong moments, there is one secondary, but much weaker, interaction between the strong and one 
induced moment.
In this case, the Curie temperature, calculated under the mean-field approximation (MFA),
seems to be enhanced due to this secondary interaction, irrespective of the sign of the secondary
interaction. In other words, the Curie temperature would be somewhat higher than that calculated by considering
only the interaction between the strong moments. The corresponding results under the random phase 
approximation (RPA) have to be obtained by solving two equations simultaneously. One can assume that the RPA results for the
Curie temperature follow the trends represented by the MFA results, being only somewhat smaller, as observed in the absence of
induced moments. In our case, since there are at least two secondary interactions (Cr-X and Cr-ES1) to consider in addition to the main
Cr-Cr interaction, the influence of these secondary interactions is definitely more complex. 

In view of the above-described situation involving secondary interactions between Cr- and the induced moments 
for the FM reference states, we have adopted the following strategy. Since no induced moments appear in calculations 
for the DLM reference states, the Curie temperature $T_c$ for these can be calculated as usual from the exchange interaction between the Cr-atoms,
i.e. the strong moments. For these cases the calculation of $T_c$ can proceed in a straightforward manner by
making use of the mean-field approximation (MFA) or the more accurate random-phase
approximation (RPA)\cite{Prange}. 
One can obtain the MFA estimate of the Curie temperature from
\begin{equation}\label{e2}
k_{B} \, T_{c}^{\rm MFA} = 
\frac{2}{3}\sum_{i \ne 0} \, J^{\rm Cr,Cr}_{0i} \, ,
\end{equation}
where the sum extends over all the neighboring shells.
An improved description of finite-temperature magnetism
is provided by the RPA, with $T_{c}$ given by 
\begin{equation}\label{e3}
(k_{B} \, T_{c}^{\rm RPA})^{-1} = \frac{3}{2} \frac{1}{N} \;
\sum_{\bf q} \, [ J^{\rm Cr,Cr}({\bf 0})-J^{\rm Cr,Cr}({\bf q}) ]^{-1} 
\, .
\end{equation}
Here $N$ denotes 
the order of the translational group applied
and $J^{\rm Cr,Cr} (\bf q)$ is the lattice Fourier 
transform of the real-space exchange integrals $J^{\rm Cr,Cr}_{ij}$.
It can be shown that $T_{c}^{\rm RPA}$ is always smaller than
$T_{c}^{\rm MFA}$ \cite{Pajda2001}. It has been shown that the RPA Curie temperatures are 
usually close to those obtained from 
Monte-Carlo simulations \cite {Rusz2006}. As shown by Sandratskii \etal\cite{Sandratskii} the calculation of $T_c$
using RPA is considerably more involved even for the case where only one secondary interaction needs to be considered,
in addition to the principal interaction between the strong moments. The complexity of the problem increases even for MFA,
if more than one secondary interaction is to be considered. The same comment applies to stability analysis using the
lattice Fourier transform of the exchange interactions. The deviation of the nature of the ground state from a collinear and parallel 
alignment of the Cr moments in the FM reference states could be studied by examining the lattice Fourier transform of the exchange 
interaction between the Cr atoms:
$J({\bf q})= \sum_{{\bf q}}J^{\rm Cr,Cr}_{0{\bf R}}\exp\left(i{\bf q}\cdot{\bf R}\right)$, if all the secondary interactions could
be ignored. This is definitely not possible for many of our FM results, where several pairs of interaction need to be considered,
and $J({\bf q})$ is a matrix bearing a complicated relationship to the energy as a function of the wave-vector {\bf q}.
Thus, in the following the results for $T_c$ will be presented mostly for the DLM reference states. For comparison, in a small number
of cases we will present $T_c$ calculated for the FM reference states using only the Cr-Cr exchange interactions as the input.
Of course, this will be done with caution only for cases where we have reason to believe that the results are at least qualitatively 
correct. Some FM results will also be included towards the stability analysis based on $J({\bf q})$ derived from Cr-Cr interactions only.
Again, this will be done with caution, only if  the corresponding results can be shown to be meaningful via additional calculations .

\subsection{Exchange interactions for the FM and DLM reference states}

\begin{figure}
\includegraphics[angle=270,width=3.95in]{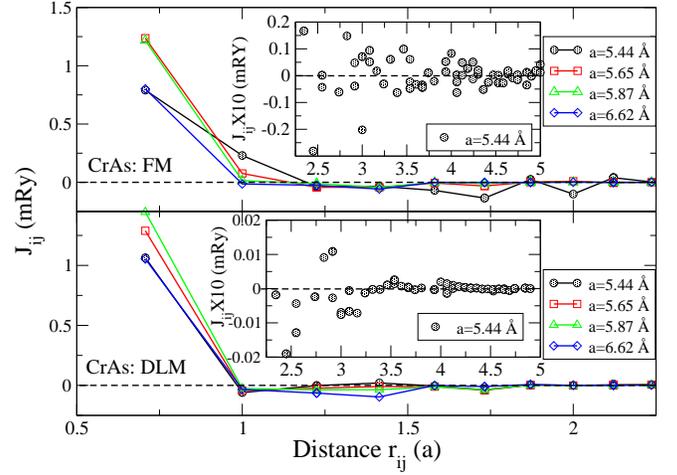}
\caption[]{(Color online) Distance dependence of the exchange interaction $J_{ij}$ between the Cr atoms in 
CrAs for various  lattice parameters $a$, calculated for the FM and DLM reference states. The distance between the Cr atoms 
is given in units of the lattice parameter $a$ (the same applies to Figs.\ref{fig11}-\ref{fig13}). 
The main plot in Fig.\ref{fig10} shows the distance dependence up to
2.25$a$, while the inset shows the values between 2.25$a$ and 5$a$. Although the individual values of $J_{ij}$ are small beyond about 2.25$a$,
their cumulative effects on the total exchange constant and the Curie temerature cannot be neglected (see text for details).
Comparison of the insets for the FM and DLM cases shows that the interactions are more damped for the DLM case, being at least an order of 
magnitude smaller for distances beyond $\sim$2.25-2.5$a$ or 15-20 neighbor shells. Similar comments apply to the interactions presented in Figs.\ref{fig11}-\ref{fig13}.}
\label{fig10}
\end{figure}

The Cr-Cr exchange interactions for all the alloys studied and for both FM and DLM reference states
become negligible as the inter-atomic distance reaches about three lattice parameters or, equivalently,
thirty neighbor shells. The same applies to the Cr-X and Cr-ES interactions for the FM cases, these interactions
in general being  somewhat smaller. The Cr-Cr interactions for the DLM reference states are more damped compared with
the corresponding FM results, showing less fluctuations in both sign and magnitude.
The distance dependence of the exchange interactions between the Cr atoms in CrAs is shown in Fig.\ref{fig10} 
for several lattice parameters. Although the nearest neighbor interaction is always positive (i.e, of 
ferromagnetic nature), the interactions with more distant neighbors are sometimes antiferromagnetic. 
Such antiferromagnetic interactions are more common in CrAs for lower lattice parameters. With increasing 
lattice parameter, interactions become predominantly ferromagnetic, and by the time the equilibrium 
lattice parameter of 5.52 \AA$\;$ is reached, antiferromagnetic interactions mostly disappear. We have 
calculated such interactions up to the 405th neighbor shell, which amounts to a distance of roughly 
8 lattice parameters. Although the interactions themselves are negligible around and after the 30$^{th}$ neighbor
shell, their influence on the lattice sums continues up to about 100 neighbor shells. By about the neighbor 110$^{th}$ shell 
(a distance of $\sim$ 5 lattice parameters) the interactions fall to values
small enough so as not to have any significant effect on the calculated lattice Fourier transform of the
exchange interaction and the Curie temperature (see below). It is clear from Fig.\ref{fig10} that ferromagnetism in 
CrAs is robust and exists over a wide range of lattice 
parameters. The distance dependence of the Cr-Cr exchange interactions in CrSb is very similar to that in CrAs for both
FM and DLM reference states.

\begin{figure}
\includegraphics[angle=270,width=3.5in]{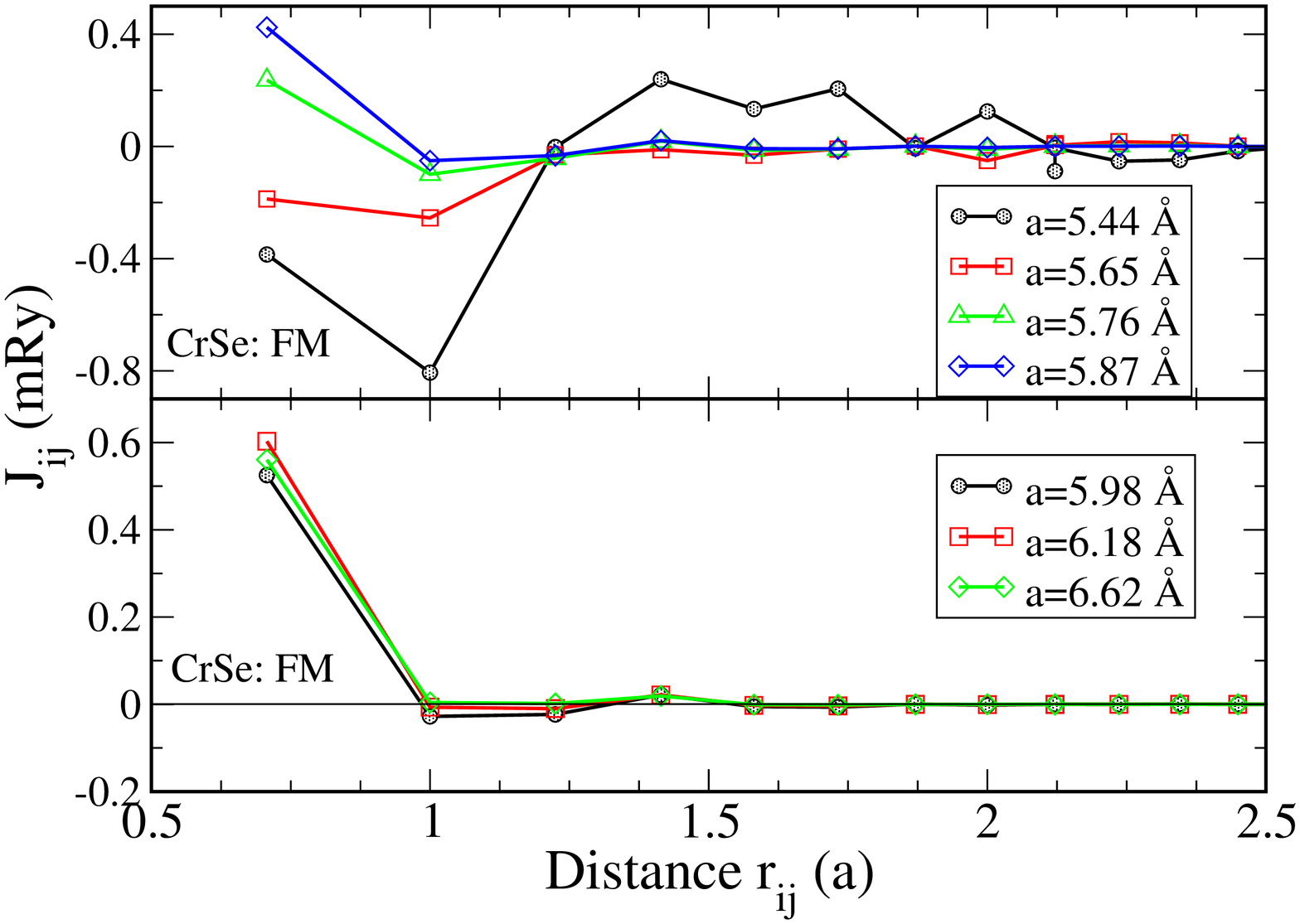}
\caption[]{(Color online) Distance dependence of the exchange interactions between the Cr atoms in CrSe for the FM reference state. 
See caption of Fig.\ref{fig10}.  }
\label{fig11}
\end{figure}

\begin{figure}
\includegraphics[angle=270,width=3.5in]{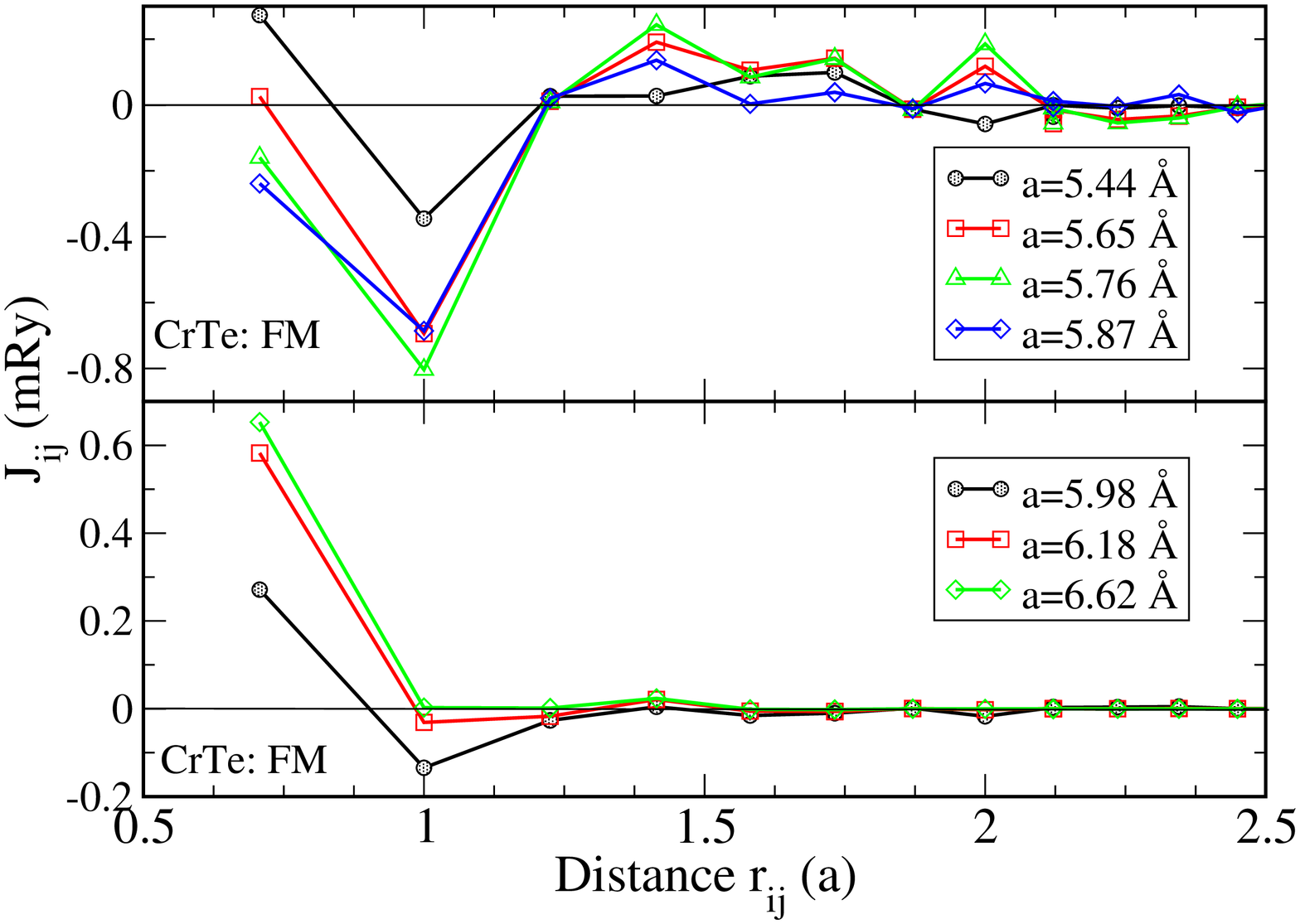}
\caption[]{(Color online) Distance dependence of the exchange interactions between the Cr atoms in CrTe for the
FM reference state. See caption of Fig.\ref{fig10}.  }
\label{fig12}
\end{figure}

\begin{figure}
\includegraphics[angle=270,width=3.5in]{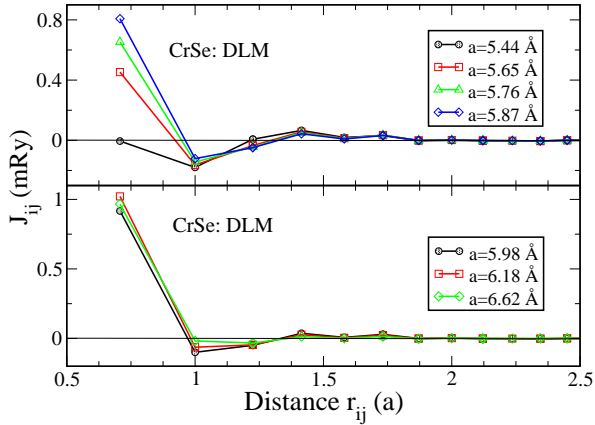}
\caption[]{(Color online) Distance dependence of the exchange interactions between the Cr atoms in CrSe for the
DLM reference state. See caption of Fig.\ref{fig10}.  }
\label{fig13}
\end{figure}

For CrS, CrSe, and CrTe the situation is somewhat different. 
For CrS and CrSe, the FM reference states for some low lattice parameters  yield Cr-Cr interactions that are antiferromagnetic
even at the nearest neighbor separation. For CrTe, at the lowest lattice parameter 
studied (5.44 \AA) the nearest neighbor interaction for the FM reference state is ferromagnetic, but becomes antiferromagnetic 
with increasing lattice parameter, changing back to ferromagnetic at higher lattice parameters. For 
all three compounds, the interactions are predominantly ferromagnetic at higher lattice 
parameters.  Figs.\ref{fig11} and \ref{fig12} show the 
distance dependence of the exchange interactions calculated for the FM reference states in CrSe and CrTe, respectively, 
for several lattice parameters. Predominant nearest neighbor antiferromagnetic interactions between the Cr atoms result in negative values
of the Curie temperature, when calculated via Eqs. (\ref{e2}) or (\ref{e3}). These results for the Curie temperature for the FM reference states can be discarded as being unphysical on two grounds: because of the neglect of the interactions involving the induced moments and also because they point to the
possibility that the ground state is most probably antiferromagnetic or of complex magnetic structure. The antiferromagnetic Cr-Cr interactions mostly disappear, when calculated for the DLM reference states. This could be interpreted as being an indication that the actual magnetic structure of the ground states for these low lattice parameters in case of CrS, CrSe and CrTe is closer to a DLM state than to an FM state. In Fig.\ref{fig13} we
show the Cr-Cr exchange interactions for the DLM reference states in case of CrSe for the same lattice parameters as those considered for Fig.\ref{fig11}. A comparison of the two figures shows that all interactions have moved towards becoming more ferromagnetic for the DLM reference states, the nearest neighbor 
interaction for the lowest lattice parameter staying marginally antiferromagnetic. 


\subsection{Stability analysis via the Lattice Fourier transform of Cr-Cr exchange interactions}
\label{subsec:stabilityJq}
The deviation of the nature of the ground state from the reference state can be studied by examining the lattice Fourier transform of the 
corresponding exchange 
interactions between the Cr atoms:
$J({\bf q})= \sum_{{\bf q}}J^{\rm Cr,Cr}_{0{\bf R}}\exp\left(i{\bf q}\cdot{\bf R}\right)$.
As pointed out earlier, for the FM reference states this procedure suffers from the drawback of neglecting the effects of all other
interactions involving the induced moments. For the DLM reference states there are no induced moments, so the relationship between the
energy and $J({\bf q})$ is simpler, but a physical picture of the spin arrangement corresponding to a particular wave-vector ${\bf q}$ is
harder to visualize. For the FM reference states, if there were no moments other than those on the Cr atoms, a maximum in $J({\bf q})$ at
${\bf q}=0$ would imply that the ground state is ferromagnetic with collinear and parallel Cr magnetic moments in all the unit cells.
A maximum at symmetry points other than the $\Gamma$-point would imply the ground state being antiferromagnetic or a spin-spiral state. 
A maximum at a wave-vector ${\bf q}$ that is not a symmetry point of the BZ would imply the ground state being an incommensurate 
spin spiral. The presence of induced moments and the consequent interactions involving non-Cr atoms and empty spheres spoil such 
interpretations based on $J({\bf q})$ derived from Cr-Cr interactions alone. However, the tendencies they reveal might still be useful. 
It is for this reason that we study the Fourier transform $J({\bf q})$, defined above, for both FM and DLM reference states.

\begin{figure}
\includegraphics[angle=270,width=3.75in]{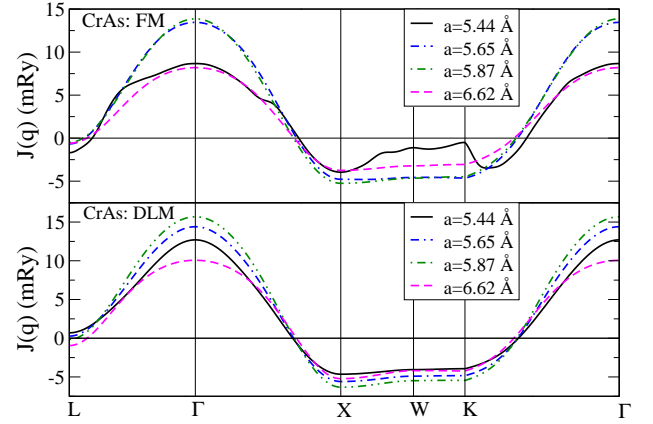}
\caption[]{(Color online) Lattice Fourier transform of the Cr-Cr exchange interactions in CrAs, calculated for the FM and DLM reference states.  }
\label{fig14}
\end{figure}

In Fig.\ref{fig14} we have plotted this quantity for CrAs. The results for CrSb are quite similar. The maximum in $J({\bf q})$ at the $\Gamma$-point for all lattice parameters and for both FM and DLM reference states can be taken as an indication that the ground state
magnetic structure is ferromagnetic for CrAs for all the lattice parameters studied.  The same comment applies to CrSb. The apparent lack of smoothness in  $J({\bf q})$ shown for the FM reference states is a consequence of the fact that there are other additional bands 
(involving induced moments), which are supposed to cross the band shown, but have not been computed.
\begin{figure}
\includegraphics[angle=270,width=3.75in]{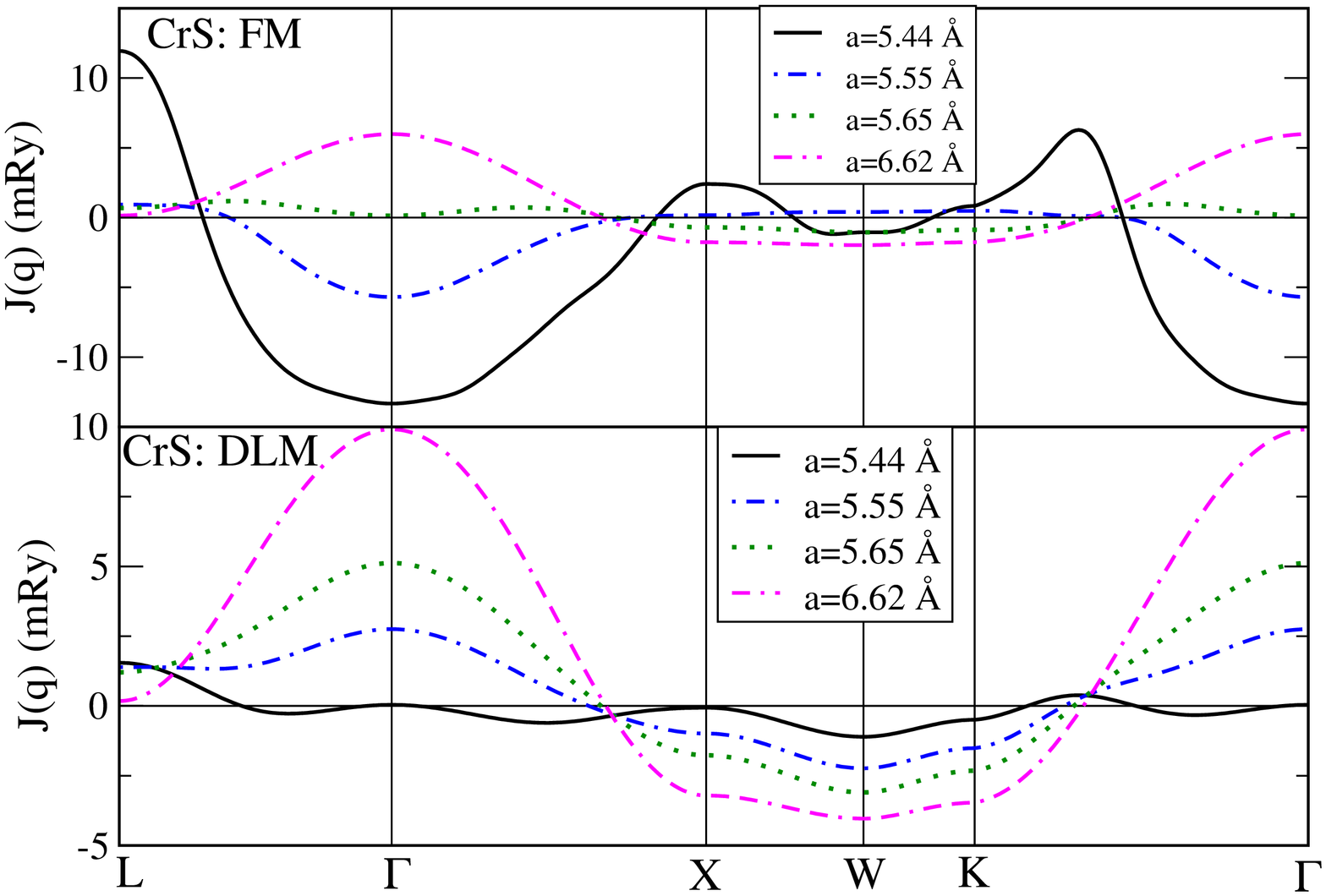}
\caption[]{(Color online) Lattice Fourier transform of the Cr-Cr exchange interactions in CrS, calculated for the FM and  DLM reference states.   }
\label{fig15}
\end{figure}
\begin{figure}
\includegraphics[angle=270,width=3.75in]{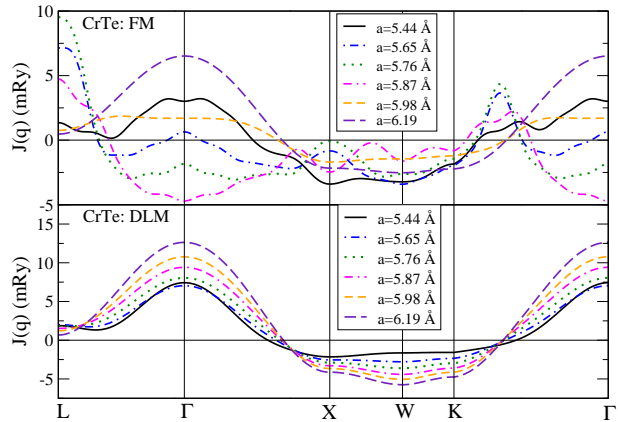}
\caption[]{(Color online) Lattice Fourier transform of the Cr-Cr exchange interactions in CrTe, calculated for the FM and  DLM reference states.   }
\label{fig16}
\end{figure}
 
 For CrS, 
CrSe, and CrTe (see Figs.\ref{fig15}-\ref{fig16}), the deviation of the ground state for low lattice parameters from the  parallel arrangement 
of Cr moments is reflected in the result that the maximum moves away from the $\Gamma$-point for the 
FM reference states. At high values of the lattice parameter the maximum returns to the $\Gamma$-point. 
The curves for CrSe are similar to those for CrS and have therefore not been shown. The fact that the maximum for the 
DLM reference sates lies at the $\Gamma$-point in most cases is again an indication that the ground state magnetic structure 
is closer to the DLM state than to the FM state. The conclusions based on the FM reference
state results in Figs.\ref{fig15} and\ref{fig16} may be suspect on ground of neglecting the interactions involving the 
induced moments. However, to explore whether they do carry any relevant information we have carried out additional calculations 
for the three compounds CrS, CrSe and CrTe for two commonly occurring antiferromagnetic configurations: AFM[001], AFM[111]. 
Note that another commonly occurring AFM configuration AFM[110] is not unique, i.e. there are several configurations that could be seen as an AFM[110] arrangement (see Fig3. of Ref.[\onlinecite{Jiang}], Table 2 of  Ref.[\onlinecite{Curtarolo}]). The simplest among 
these is actually equivalent to AFM[100]. The results for the total energy for the two AFM calculations
are shown in Table \ref{table2} and compared with the corresponding FM and DLM total energies. For CrS, the 
lowest energy state for lattice parameters 5.44 and 5.55 \AA$\;$ is AFM[111], exactly as suggested by the maximum 
in $J({\bf q})$ appearing at the L-point in Fig.\ref{fig15} for the
FM reference state and for these two lattice parameters. As the lattice parameter increases beyond 5.55 \AA, antiferromagnetic interactions diminish. For the next higher lattice parameter 5.66 \AA$\;$in Table \ref{table2}, the lowest energy state is DLM. This may suggest that the
ground state has a complex magnetic structure, which remains to be explored. For higher lattice parameters the FM state has the lowest energy.
For CrSe, AFM[111] state has the lowest energy up to the lattice parameter 5.66 \AA, as is also supported by the maximum of $J({\bf q})$ at L-point. The $J({\bf q})$ curves for CrSe are similar to those of CrS, and have not been shown. For CrTe, at the lowest lattice parameter of 5.44 
\AA$\;$ the lowest energy state is FM, as is also indicated by the maximum of $J({\bf q})$ at the $\Gamma$-point. For higher lattice parameters 5.65 and 5.76\AA, even though the $J({\bf q})$ curves point to the possibility of an AFM[111] ground state, the FM state energy turns out to be the lowest among the configurations studied. It could be concluded that in this case a proper relationship between the energy and $J({\bf q})$, obtained without the neglect of the induced moments, would point to the ground state being FM. For these three chalcogenides, for lattice
parameters above 5.65-5.7 \AA$\;$the ground state should be FM. 
\begin{table*}
\caption{Comparison of total energies per atom (Ry) in the FM, DLM, AFM[001], AFM[111], and AFM[110] states as a function of the lattice parameter 
for CrS, CrSe, and CrTe. Results for 5 lattice parameter values are shown, usually to 4 places after the decimal, 5 only to break a tie.}
\label{table2}
\begin{ruledtabular}
\begin{tabular}{lcccccc}
Lattice parameter (\AA) & 5.44 & 5.55 & 5.66 & 5.76 & 5.87  \\
{\bf CrS}\\
DLM  & -723.4504 & -723.4468 & -723.4433 & -723.4381 & -723.4332\\
FM  & -723.4492 &  -723.4465 & -723.4432 & -723.4395 & -723.4351\\
AFM[001] & -732.4503 & -723.4464 & -723.4420 &-723.4375 &-723.4325\\
AFM[111] & -723.4506 & -723.4469 & -723.4427 &-723.4382 &-723.4333\\
{\bf CrSe}\\ 
DLM  &  -1737.91456 & -1737.91388 & -1737.9123 &-1737.9100 &  -1737.9071\\
FM  & -1737.9139 & -1737.9132 & -1737.91236 & -1737.9110 & -1737.9087  \\
AFM[001] & -1737.9144 & -1737.9136 & -1737.9118 &-1737.9094 &-1737.9065\\
AFM[111] & -1737.91458 & -1737.91398 & -1737.91239 &-1737.9101 &-1737.9071\\
{\bf CrTe}\\ 
DLM  & -3918.9074 & -3918.9124 & -3918.9158 & -3918.9159 & -3918.9189  \\
FM  & -3918.9078 & -3918.9128 & -3918.9162 & -3918.9181 & -3918.9194 \\
AFM[001] & -3918.9170 & -3918.9120 & -3918.9154 &-3918.9174 &-3918.9182\\
AFM[111] & -3918.9072 & -3918.9123 & -3918.9159 &-3918.9180 &-3918.9189 \\
\end{tabular}
\end{ruledtabular}
\end{table*}
\subsection{Curie temperatures}
\begin{figure}
\includegraphics[angle=270,width=3.75in]{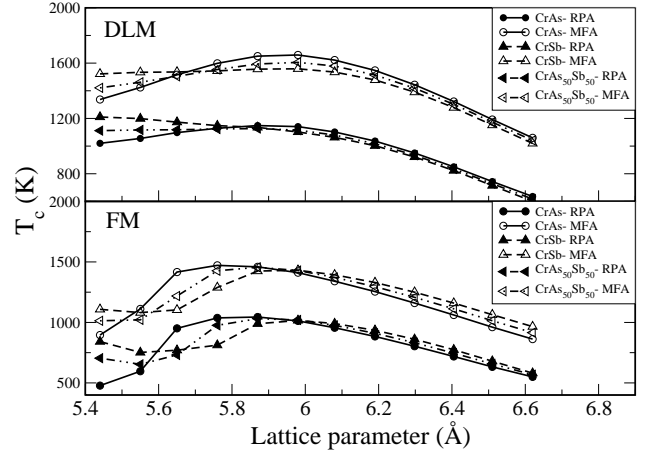}
\caption[]{Curie temperatures in ZB CrAs and CrSb compounds for FM and DLM reference states. For comparison, the results for the mixed
alloy CrAs$_{50}$Sb$_{50}$ are also shown.  }
\label{fig17}
\end{figure}

We determine the Curie temperature using Eqs. (\ref{e2}) and (\ref{e3}). For the DLM reference states, these produce estimates of
$T_c$ from above the ferromagnetic$\leftrightarrow$paramagnetic transition, and are free from errors due to induced moments. However,
these estimates are high compared with properly derived values of $T_c$ from below the transition. The latter estimates would require the
use of FM reference states (where the ground states are known to be FM) and thus a proper treatment of the induced moments. For CrAs and CrSb,
the magnetic state is ferromagnetic for all the lattice parameters considered. Hence, for the sake of comparison we have calculated the
$T_c$ for the FM reference states using Eqs. (\ref{e2}) and (\ref{e3}) as well. According to the results of Sandratskii \etal\cite{Sandratskii}
the correctly calculated $T_c$ values, in the presence of  interactions involving all the induced moments, would be higher. Thus, the correct
estimates of $T_c$ should lie somewhere between the DLM results and the FM results obtained with the neglect of the induced moments. In 
Fig. \ref{fig17} we show these results for CrAs, CrSb. We have used up to 111 shells in the evaluation of Eq. (\ref{e2}) 
and for the lattice Fourier transform of $J^{\rm Cr,Cr}({\bf q})$ 
in Eq.~(\ref{e3}), after having tested the convergence with respect to the 
number of shells included.
The estimated computational error corresponding to the chosen
number of shells used in these calculations is below $\pm 2~K$.
For comparison we also include the results for the mixed alloy 
CrAs$_{50}$Sb$_{50}$, for which the calculated $T_c$ values fall, as expected, in between those of
CrAs and CrSb. Since RPA values are more accurate than MFA values, our best estimates of $T_c$ for CrAs range from somewhat higher than 500 K
at low values of the lattice parameter, increasing to 1000-1100 K around the mid lattice parameter range (5.75-5.9 \AA) and then decreasing to
around 600 K for higher lattice parameters (6.5 \AA$\;$ and above). For CrSb these estimates are consistently higher than those for CrAs: 1100 K, 1500 K and 1200 K, respectively. The estimates for CrAs are similar to those provided by Sasaiolglu \etal \cite{Sasaiolglu} 

\begin{figure}
\includegraphics[angle=270,width=3.5in]{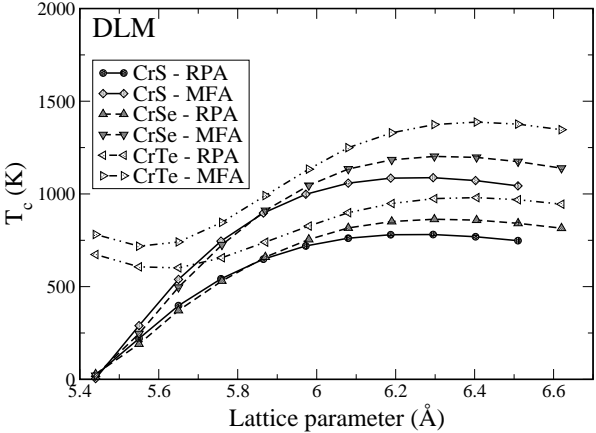}
\caption[]{Curie temperatures in ZB CrS, CrSe and CrTe calculated for DLM reference states. For CrS and CrSe the ground state is
antiferromagnetic (see text for details) for some low values of the lattice parameter. The values shown for those lattice parameters should
be discarded. }
\label{fig18}
\end{figure}

For CrS, CrSe,  the results obtained with the FM reference states would clearly be wrong, in particular, for the low values
of the lattice parameters, for which we have shown the ground state to be antiferromagnetic within our limited search. There is a possibility that
the ground state for certain lattice parameters might have a complex magnetic structure. For CrTe, even though the ground state appears to be
ferromagnetic, there are considerable antiferromagnetic spin fluctuations, making the FM estimates unreliable. In Fig. \ref{fig18} we show the $T_c$ values for
CrS, CrSe and CrTe for the DLM reference states. The values for lattice parameters for which the ground state has been shown to be antiferromagnetic in the preceding section should be discarded as being inapplicable. 

\begin{figure}
\includegraphics[angle=270,width=3.5in]{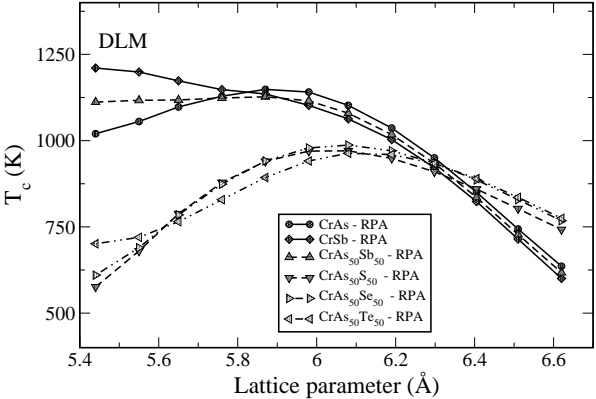}
\caption[]{Variation of Curie temperature as a function of lattice parameters in ZB 
CrAs$_{50}$X$_{50}$ alloys with X=S, Se and Te. For comparison, the results for CrAs, CrSb and 
CrAs$_{50}$Sb$_{50}$ are also shown.  All results shown are for DLM reference states, and as such, should be considered as upper limits for $T_c$. }
\label{fig19}
\end{figure}

Similar results for the alloys CrAs$_{50}$X$_{50}$ (X=S, Se and Te) are shown in Fig. \ref{fig19} for DLM reference states. For these, the ground state is ferromagnetic for all lattice parameters. However, because of the neglect of the induced moments related effects, our results for the
Curie temperatures for the FM reference states are lower than the properly calculated values. Thus in Fig.\ref{fig19}  we show the DLM results only, which are devoid of the induced moment effects and 
 provide us with estimates of $T_c$  from above the transition. These  are expected to be somewhat higher than the properly computed values for FM reference states.
Thus, for these alloys the trend revealed  in Fig. \ref{fig19} for the variation of $T_c$ with lattice parameter is correct.  The estimates themselves are qualitatively correct, albeit somewhat higher than the correct values. Only the RPA values are plotted in Fig. \ref{fig19}, which are more reliable than the MFA values. For comparison, we also show the results for the pnictides CrAs, CrSb, and  CrAs$_{50}$Sb$_{50}$, which are isoelectronic among themselves, but have half an electron per unit cell less than the mixed alloys CrAs$_{50}$X$_{50}$ (X=S, Se and Te) 

The differences between the results for the pnictides, chalcogenides and the mixed pnictide-chalcogenides can be summarized as follows. The pnictides, CrAs, CrSb, and  CrAs$_{50}$Sb$_{50}$, are strong ferromagnets at all the lattice parameters studied (5.44 \AA - 6.62 \AA). In the DLM description, their $T_c$ stays more or less constant (apart from a minor increase) as the lattice parameter increases from 5.4 /AA to 6.1 \AA, and then decreases beyond (Figs. \ref{fig17} and \ref{fig19}. The chalcogenides are antiferromagnetic or have complex magnetic structure for low lattice parameters. In the DLM description, their $T_c$ in the ferromagnetic state increases and then becomes more or less constant as the lattice parameter increases (Fig. \ref{fig18}). The mixed alloys CrAs$_{50}$X$_{50}$ (X=S, Se, Te) are ferromagnetic at all the lattice parameters studied.
In the DLM description, their $T_c$ rises and then falls as the lattice parameter is increased from 5.44 \AA to 6.62 \AA. A comparison of the results presented in Figs. \ref{fig17}-\ref{fig19} shows that large changes in $T_c$ take place by changing the number of carriers. Changes due to
isoelectronic doping are small compared with changes brought about by changing carrier concentration.

\section{Summary of results}
Our \textit{ab initio} studies of the electronic structure, magnetic moments, exchange 
interactions and Curie temperatures in ZB CrX (X=As, Sb, S, Se and Te) and 
CrAs$_{50}$X$_{50}$ (X=Sb, S, Se and Te) reveal that half-metallicity in these alloys 
is maintained over a wide range of lattice parameters. The results for the exchange 
interaction and the Curie temperature show that these alloys have relatively high Curie 
temperatures, i.e. room temperature and above. The exceptions occur for the alloys involving 
S, Se and Te at some low values of lattice parameters, where significant inter-atomic 
antiferromagnetic exchange interactions indicate ground states to be either antiferromagnetic 
or of complex magnetic nature. A comparison of total energies for the FM, DLM, and two ZB antiferromagnetic
configurations (AFM[001] and AFM[111]) show the lowest energy configuration to be AFM[111] for CrS and CrSe for compressed lattice parameters (Table\ref{table2}). The possibility of AFM ground states for compressed lattice parameters for CrS was noted by
Zhao and Zunger\cite{Zunger2005} and for CrSe by Sasioglu \etal \cite{Sasaiolglu}. Our search for the antiferromagnetic ground states is
more thorough than what was reported in these two studies. An extensive study of several antiferromagnetic configurations 
as well as ferrimagnetic
and more complex magnetic structures for CrS, CrSe and CrTe is currently underway.

The mixed pnictide-chalcogenide alloys CrAs$_{50}$X$_{50}$ (X= S, Se, Te) do not
show any tendency to antiferromagnetic spin fluctuations for the entire range of the lattice parameter studied. Presumably the
pnictogens suppress antiferromagnetic tendencies. Such alloys may play an important role in fabricating stable ZB half-metallic materials, as the
concentration of the pnictogens and the chalcogens may be varied to achieve lattice-matching with a given substrate. As long as the concentration of As or Sb is higher than the chalcogen concentration,  half-metallic ferromagnetic state can be achieved. There is a large variation in the Curie
temperature of these alloys (Fig. \ref{fig19}) as the lattice parameter varies from the low ($\sim$ 5.4 \AA) to the mid ($\sim$ 6.1 \AA) range of the lattice parameters studied. This variation is much smaller for the isoelectronic alloys CrAs, CrSb and CrAs$_{50}$Sb$_{50}$ (Fig. \ref{fig17}) over this range of lattice parameters. Note that most II-VI and III-V ZB semiconductors have lattice parameters in this range. Large changes in $T_c$ can be brought about by changing the carrier concentrations. The pnictides in general have a higher $T_c$ than the chalcogenides.

Our results for the Curie temperature, the lattice Fourier transform
of the exchange interactions,
and the resulting stability analysis are based on the exchange interactions between the Cr 
atoms only. For the FM reference states this causes some errors due to the neglect of the effects of the induced moments.
The DLM results are free from such errors. It is expected that the present study will provide both  qualitative and quantitative
guidance to experimentalists in the field.

ACKNOWLEDGMENTS 

This work was supported by a grant from the Natural Sciences and Engineering Research Council of Canada.


\begin{thebibliography}{99}      
\bibitem{Sato1} K. Sato and H. Katayama-Yoshida, Semicond. Sci. Technol. {\bf 17}, 367 (2002).
\bibitem{Sato4} See K. Sato, T. Fukushima and H. Katayama-Yoshida, J. Phys.: Condens. Matter {\bf 19}, 365212 (2007), and references therein.
\bibitem{Sato5} See B. Belhadji, L. Bergqvist, R. Zeller, P.H. Dederichs, K. Sato and H. Katayama-Yoshida, J. Phys.: Condens. Matter {\bf 19}, 436227 (2007), and references therein.  
\bibitem{Saito} H. Saito, V. Zayets, S. Yamagata, and K. Ando, Phys. Rev. Lett {\bf 90}, 207202-1 (2003).
\bibitem{Sato6} K. Sato and H. Katayama-Yoshida, Jpn. J. Appl. Phys. {\bf 40}, L651 (2001).
\bibitem{Akinaga2000} H. Akinaga, T. Manago, and M. Shirai, Jpn. J. Appl. Phys.{\bf 39}, L1118 (2000).
\bibitem{Li2008} S. Li, J-G Duh, F. Bao, K-X Liu, C-L Kuo, X. Wu, Liya L\"{u}, Z. Huang, and Y Du,
J. Phys. D: Appl. Phys. {\bf 41} 175004 (2008).
\bibitem{Shirai2003} M. Shirai, J. Appl. Phys. {\bf 93}, 6844 (2003).
\bibitem{Akinaga2005} H. Akinaga, M. Mizuguchi, K. Nagao, Y. Miura, and M. Shirai in {\it Springer Lecture Notes in Physics} {\bf 676},
293-311 (Springer-Verlag, Berlin 2005).
\bibitem{Yamana2004} K. Yamana, M. Geshi, H. Tsukamoto, I. Uchida, M. Shirai, K. Kusakabe, and N. Suzuki,
 J. Phys.: Condens. Matter {\bf 16}, S5815 (2004).
\bibitem{Kahal2007} L. Kahal, A. Zaoul, M. Ferhat, J. Appl. Phys. {\bf 101}, 093912 (2007).
\bibitem{Galanakis2003} I. Galanakis and P. Mavropoulos, \prb {\bf 67}, 104417 (2003); see also I. Galanakis, \prb {\bf 66}, 012406 (2002).  
\bibitem{Pask} J.E. Pask, L.H. yang, C.Y. Fong, W.E. Pickett, and S. Dag, \prb {\bf 67}, 224420 (2003).
\bibitem{Ito} T. Ito, H. Ido, and K. Motizuki, J. Mag. Mag. Mat. {\bf 310}, e558 (2007).
\bibitem{Shi} L-J Shi and B-G Liu, J. Phys.: Condens. Matter {\bf 17}, 1209 (2005). 
\bibitem{Zhang} M. Zhang \etal , J. Phys.: Condens. Matter {\bf 15}, 5017 (2003).
\bibitem{Kubler2003} J. K\"{u}bler, \prb {\bf 67}, 220403(R) (2003).
\bibitem{Sanyal2003} B. Sanyal, L. Bergqvist, and O. Eriksson, \prb {\bf 68}, 054417 (2003).
\bibitem{Pettifor2003} W-H. Xie, Y-Q. Xu, B-G. Liu, and D.G. Pettifor  \pl {\bf 91}, 037204 (2003).
\bibitem{Zhao2001} J.H. Zhao, F. Matsukura, K. Takamura, E. Abe, D. Chiba, and H. Ohno, Appl. Phys. Lett.
{\bf 79}, 2776 (2001).
\bibitem{Ono2001} K. Ono, J. Okabayashi, M. Mizuguchi, M. Oshima, A. Fujimori, and H. Akinaga, J. Appl. Phys. {\bf 91}, 8088 (2001).
\bibitem{Zunger2005} Y-J. Zhao and A. Zunger, \prb {\bf 71}, 132403 (2005).
\bibitem{Deng2006} J.J. Deng, J.H. Zhao, J.F. Bi, Z.C. Niu, F.H. Yang, X.G. Wu, and H.Z. Zheng, J. Appl. Phys. {\bf 99},
093902 (2006).
\bibitem{Kudrnovsky1990} J. Kudrnovsk\'y and V. Drchal, \prb {\bf 41}, 7515 (1990).
\bibitem{Turek97} I. Turek, V. Drchal, J. Kudrnovsk\'y, M. \v{S}ob, and
P. Weinberger, {\it Electronic Structure of Disordered Alloys,
Surfaces and Interfaces} (Kluwer, Boston-London-Dordrecht, 1997).
\bibitem{VWN} S.H. Vosko, L. Wilk, and M. Nusair,
Can. J. Phys. {\bf 58}, 1200 (1980).
\bibitem{savrasov} S.Yu. Savrasov, and D.Yu. Savrasov, \prb {\bf 46}, 12181 (1992).
\bibitem{Heine1} V. Heine, J.H. Samson, and C.M.M. Nex, J. Phys. F: Met. Phys. {\bf 11}, 2645 (1981).
\bibitem{Heine2} V. Heine and J.H. Samson, J. Phys. F: Met. Phys. {\bf 13}, 2155 (1983).
\bibitem{Hasegawa} H. Hasegawa, J. Phys. Soc. Jpn. {\bf 46}, 1504 (1979).
\bibitem{Pettifor} D.G. Pettifor, J. Magn. Magn. Mater {\bf 15-18}, 847 (1980).  
\bibitem{Staunton} J.B. Staunton, B.L. Gyorffy, A.J. Pindor, G.M. Stocks, and H. Winter,
J. Phys. F {\bf 15}, 1387 (1985).
\bibitem{Pindor} A.J. Pindor, J. Staunton, G.M. Stocks, H. Winter, J. Phys. F {\bf 13}, 979 (1983).
\bibitem{Sasaiolglu} E. Sasaio$\tilde{g}$lu, I. Galanakis, L.M. Sandratskii, and P. Bruno,  J. Phys: Condens. Matter
 {\bf 17} 3915 (2005).
\bibitem{Sandratskii} L.M. Sandratskii, R. Singer, and E. Sasio\u{g}lu, \prb {\bf 76}, 184406 (2007).
\bibitem{Pajda2001} M. Pajda, J. Kudrnovsk\'y, I. Turek, V. Drchal,
and P. Bruno, Phys. Rev. B {\bf 64}, 174402 (2001).
\bibitem{Liechtenstein84I} A.I. Liechtenstein, M.I. Katsnelson and
V.A. Gubanov, J. Phys.F: Met.Phys. {\bf14}, L125 (1984).
\bibitem{Liechtenstein87} A. I. Liechtenstein, M. I. Katsnelson, V. P. Antropov, V. A. Gubanov, J. Magn. Magn. Mater.
{\bf 67}, 65 (1987).
\bibitem{Liechtenstein84II} A.I. Liechtenstein, M.I. Katsnelson and
V.A. Gubanov, Solid.State.Commun. {\bf 51}, 1232 (1984).
\bibitem{Liechtenstein88} A.I. Liechtenstein, M.I. Katsnelson, V.P. Antropov and
V.A. Gubanov, J.Magn.Magn.Mater. {\bf 21}, 35 (1988).
\bibitem{Gubanov92} V.A. Gubanov, A.I. Liechtenstein, A.V. Postnikov
{\em Magnetism and the electronic structure of crystals}, edited by
M. Cardona, P. Fulde, K. von Klitzing, H.-J. Queisser (Springer, Berlin, 1992).
\bibitem{Andersen} see, e.g., O.K. Andersen, O. Jepsen, and D. Gl\"{o}tzel, in
{\it Highlights of Condensed Matter Theory}, edited by F. Bassani \etal (North-Holland, Amsterdam,
1985), p.59.
\bibitem{Heine} V. Heine, {\it Solid State Physics} {\bf 35} (Academic Press, New York), 1 (1980).
\bibitem{Oswald} A. Oswald \etal, J. Phys. F {\bf 15}, 193 (1985).
\bibitem{Ducastelle} F. Ducastelle, ``Order and phase stability in alloys'' (North-Holland, Amsterdam,
1991).
\bibitem{Sluiter} M. Sluiter, and P.E. A. Turchi, Phys. Rev. B {\bf 40}, 11215 (1989).
\bibitem{C-W} J.W.D. Connolly and A.R. Williams, Phys. Rev. B {\bf 27}, 5169 (1983).
\bibitem{Lu} Z.W. Lu, S.-H. Wei, A. Zunger, S. Frota-Pessoa, and L.G. Ferreira, Phys. Rev. B {\bf 44},
512 (1991).
\bibitem{Sandratskii91} L.M. Sandratski, J. Phys.: Condens. Matter 3, 8565 (1991).
\bibitem{Turek2006} I. Turek, J. Kudrnovsk\'y, V. Drchal,  and P. Bruno,
Philos. Mag. {\bf 86}, 1713 (2006).
\bibitem{Prange} C.S. Wang, R.E. Prange, and V. Korenman,  Phys. Rev. B {\bf 25}, 5766 (1982).
\bibitem{Rusz2006} J. Rusz, L. Bergqvist, J. Kudrnovsk\'y, and
I. Turek, Phys. Rev. B {\bf 73}, 214412 (2006).
\bibitem{Jiang} L. Jiang, Q. Feng, Y. Yang, Z. Chen, and Z. Huang, Sol. St. Comm. {\bf 139}, 40 (2006).
\bibitem{Curtarolo} S. Curtarolo, D. Morgan and G. Ceder, Comp. Coupl. Phase Diagrams and Thermochemistry {\bf 29}, 163 (2005).
\end {thebibliography}
\end{document}